\documentclass[%
 reprint,
superscriptaddress,
 amsmath,amssymb,
 aps,
prl,
]{revtex4-2}

\usepackage{graphicx}
\usepackage{dcolumn}
\usepackage{bm}
\usepackage{xcolor}

\usepackage{soul}
\usepackage{url}
\usepackage{siunitx}

\begin{document}

\newcommand{\bra}[1]{\left\langle\,#1\,\right|}
\newcommand{\ket}[1]{\left|\,#1\,\right\rangle}

\preprint{APS/123-QED}

\title{Fermiology and spin polarization of topological surface states in PtBi$_2$}

\author {Anders~Christian~Mathisen}
\email{anders.c.mathisen@ntnu.no}
\affiliation {Center for Quantum Spintronics, Department of Physics, Norwegian University of Science and Technology, 7491 Trondheim, Norway}

\author {Xin~Liang~Tan}
\affiliation {Center for Quantum Spintronics, Department of Physics, Norwegian University of Science and Technology, 7491 Trondheim, Norway}

\author {Stefanie~Suzanne~Brinkman}
\affiliation {Center for Quantum Spintronics, Department of Physics, Norwegian University of Science and Technology, 7491 Trondheim, Norway}

\author{Kristian M{\ae}land}
\affiliation{Institute for Theoretical Physics and Astrophysics, University of W{\"u}rzburg, D-97074 W{\"u}rzburg, Germany}
\affiliation{W{\"u}rzburg-Dresden Cluster of Excellence ctd.qmat, D-97074 W{\"u}rzburg, Germany}

\author {Fabian G\"ohler}
\affiliation {Center for Quantum Spintronics, Department of Physics, Norwegian University of Science and Technology, 7491 Trondheim, Norway}

\author {\O yvind~Finnseth}
\affiliation {Department of Materials Science and Engineering, Norwegian University of Science and Technology, 7491 Trondheim, Norway}

\author{Grigory Shipunov}
\affiliation{Van der Waals-Zeeman Institute, Department of Physics and Astronomy,
University of Amsterdam, Science Park 904, 1098 XH Amsterdam, The Netherlands}

\author{Falk Pabst}
\affiliation{Van der Waals-Zeeman Institute, Department of Physics and Astronomy,
University of Amsterdam, Science Park 904, 1098 XH Amsterdam, The Netherlands}

\author{Manuel Alonso Lemos}
\affiliation{Instituto Balseiro, Univ. Nacional de Cuyo, Av. Bustillo 9500, Argentina}

\author {Balasubramanian Thiagarajan}
\affiliation {MAX IV Laboratory, Lund University, Lund, Sweden}

\author {Craig Polley}
\affiliation {MAX IV Laboratory, Lund University, Lund, Sweden}

\author{Bj{\"o}rn Trauzettel}
\affiliation{Institute for Theoretical Physics and Astrophysics, University of W{\"u}rzburg, D-97074 W{\"u}rzburg, Germany}
\affiliation{W{\"u}rzburg-Dresden Cluster of Excellence ctd.qmat, D-97074 W{\"u}rzburg, Germany}

\author{Anna Isaeva}
\affiliation{Faculty of Physics, Technical University of Dortmund, Otto-Hahn-Straße 4, D-44227 Dortmund, Germany}
\affiliation{Center Future Energy Materials and Systems (RC FEMS), Germany}\affiliation{Van der Waals-Zeeman Institute, Department of Physics and Astronomy,
University of Amsterdam, Science Park 904, 1098 XH Amsterdam, The Netherlands}

\author{Jorge I. Facio}
\affiliation{Instituto Balseiro, Univ. Nacional de Cuyo, Av. Bustillo 9500, Argentina}
\affiliation{Centro Atómico Bariloche, Instituto de Nanociencia y Nanotecnología (CNEA-CONICET), Av. Bustillo 9500, Argentina}

\author {Hendrik~Bentmann}
\affiliation {Center for Quantum Spintronics, Department of Physics, Norwegian University of Science and Technology, 7491 Trondheim, Norway}

\date{\today}

\begin{abstract} 

Layered PtBi$_2$ is a candidate for topological superconductivity arising in Fermi-arc surface states. Using spin- and angle-resolved photoemission spectroscopy, we demonstrate that the Fermi arcs in PtBi$_2$ are singly degenerate and spin-polarized, which establishes their nontrivial topology and constitutes a necessary condition for topological superconductivity. We further uncover a pronounced surface-termination dependence of the Fermi-arc dispersion, yielding either nearly flat or approximately linear bands in agreement with first-principles calculations. Together, the observed spin polarization and termination-dependent bandwidth of the Fermi-arc surface states identify key ingredients underlying the potential emergence of topological superconductivity in PtBi$_2$. 
\end{abstract}

\maketitle 

\textit{Introduction--} The interplay between many-body interactions 
and electronic band topology holds promise for the discovery of novel quantum states of matter \cite{keimer2017physics,frolov2020topological,bernevig2022progress,di2025kagome}. An important example is topological superconductivity, where a superconducting gap with nontrivial topology gives rise to edge or surface states that are expected to host Majorana quasiparticles with non-Abelian exchange statistics \cite{RevModPhys.80.1083,SatoAndo2017}.
Beyond their fundamental significance, such states could become a basis for applications in quantum technology. One proposed route to topological superconductivity exploits spin momentum textures in topological band structures \cite{FuKane:07,KobayashiSato2015,Lee:24}. 
In such systems, the spin momentum texture of the Bloch wave functions may induce unconventional or topological superconducting gaps with anisotropy and phase winding inherited from the band geometry. 
However, experimental realizations remain limited to a few candidate materials \cite{mandal2023topological}.

Recent experiments show evidence for unconventional superconductivity in the Fermi-arc surface states of the Weyl semimetal PtBi$_2$, based on angle-resolved photoelectron spectroscopy (ARPES) \cite{kuibarov_evidence_2024,changdar_topological_2025} and scanning tunneling microscopy (STM) \cite{schimmel_surface_2024,huang_sizable_2025}. ARPES measurements indicate a nodal structure of the superconducting gap consistent with an unprecedented exotic $i$-wave pairing symmetry \cite{changdar_topological_2025}. An analysis of the superconducting-gap spectra in STM indicates an anisotropic chiral pairing symmetry \cite{huang_sizable_2025}. 
Despite these encouraging findings, the nature and origin of the surface superconductivity in PtBi$_2$ remain to be firmly established, which has attracted considerable interest \cite{Nomani2023FermiArcSCDOS, Bai2025WSMSC, Trama2024TRSWeylSM_SC, Waje2025PtBi2GL, Maeland_2025_Phonon_mediated_top_supcon, maeland2025mechanismnodaltopologicalsuperconductivity,  Huang2025WeylIISC, Buccheri2026ph, Dsouza2026KL}. 
Recent theory predicts that $i$-wave pairing may arise in PtBi$_2$ from a combination of electron-phonon coupling and statically screened Coulomb interaction but its emergence depends sensitively on the bandwidth and low-energy dispersion of the surface states \cite{maeland2025mechanismnodaltopologicalsuperconductivity}. The potentially non-trivial topology of the superconducting state critically relies on the single-band nature and momentum-dependent spin texture of the Fermi-arc surface states  \cite{Maeland_2025_Phonon_mediated_top_supcon,FuKane:07}, which so far, however, has not been experimentally demonstrated. Hence, a detailed investigation of the Fermi arcs in PtBi$_2$ for their band dispersion, momentum-space topology, and spin texture becomes essential.    

In this work, we present a systematic study of the low-energy electronic structure of PtBi$_2$ in the normal state, based on a careful comparison of results from spin- and angle-resolved photoelectron spectroscopy (spin-ARPES) and first-principles calculations. Remarkably, we find that the Fermi-arc band dispersion depends strongly on the atomic surface termination of PtBi$_2$(0001). While one termination hosts a flat Fermi arc on the few-meV energy scale, the other exhibits a steep, linear dispersion extending over several hundred meV, in close agreement between experiment and theory. Our spin-resolved measurements demonstrate non-degeneracy and a helical spin texture of the Fermi arcs for both surface terminations. These findings demonstrate the topological nature of the Fermi arcs and establish prerequisites for topological superconductivity.

\begin{figure}[t!]
    \centering
    \includegraphics[width=\linewidth]{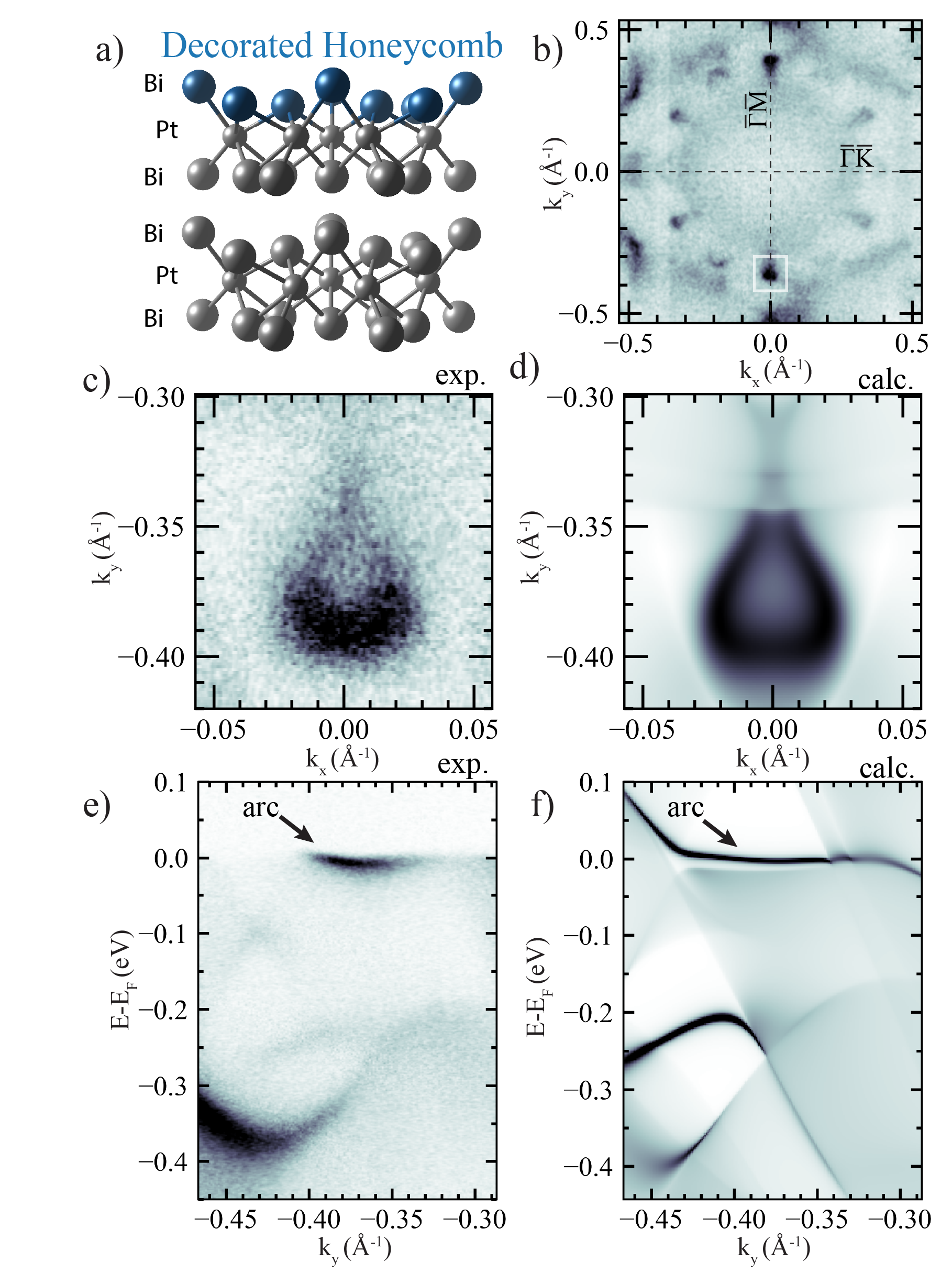}
    \caption{(a) The atomic configuration of PtBi$_2$(0001) with DH surface termination highlighted in blue. (b) Experimental Fermi surface of the DH termination from ARPES at a photon energy $h\nu =$~30 eV. The white rectangle indicates the position of one of the 6 Fermi arcs for this termination. (c) High-resolution ARPES momentum map of the Fermi arc highlighted in b). (d) Surface spectral weight calculation of the DH Fermi arc. (e) ARPES data set along $\bar{\Gamma}\bar{M}$ through the Fermi arc with corresponding surface spectral weight calculation in (f).}
    \label{figure1}
\end{figure}

\textit{Methods--} The experiments were carried out at the BLOCH beamline of the MAX IV facility (Lund, Sweden). We performed high-resolution ARPES at the A-branch endstation and spin-ARPES at the B-branch endstation of the beamline. Measurements were done in ultrahigh vacuum (UHV) below $10^{-10}$ mbar. The trigonal polymorph of PtBi$_2$ crystallizes in the space group \textit{P}31\textit{m} with $a=b$ = 6.573 Å and $c$ = 6.162 Å in the hexagonal unit cell and has broken inversion symmetry \cite{Kaiser_PtBi2_Low_T_reduction_2014,feng_rashba-like_2019,Shipunov_polymorphic_PtBi2_2020, veyrat_berezinskiikosterlitzthouless_2023}. Single crystals were grown from an overstoichiometric bismuth flux following the protocol described in Ref. \cite{Shipunov_polymorphic_PtBi2_2020}. Before measurement, the single crystals were cleaved \textit{in-situ} using kapton tape. Two distinct surface terminations may present themselves upon cleaving \cite{Wenxiang_2020_electronic_structure_of_PtBi2,kuibarov_evidence_2024}, as illustrated in Fig.~1(a) and Fig.~2(a). Following the established nomenclature in \cite{vocaturo_electronic_2024}, which builds on previous studies of PtBi$_2$ \cite{BISWAS_1969_PtBi2,Kaiser_PtBi2_Low_T_reduction_2014,Yang_Giant_linear_magneto_resistance_2016, Yao_Bulk_and_Surface_2016, Thirupathaiah_Origin_of_MR_2018, gao_tripple_degenerate_2018, feng_rashba-like_2019, Shipunov_polymorphic_PtBi2_2020, Wenxiang_2020_electronic_structure_of_PtBi2, Wang_2021_Press_Induced_SupCon, Bashlakov_2022_El_Ph_SupCon,veyrat_berezinskiikosterlitzthouless_2023}, we refer to these terminations as decorated-honeycomb (DH) and kagome-like (KL). For details on the (spin)ARPES measurements and crystal growth, see the Supplementary Material \cite{suppl}.

Based on the experimental crystal structure, we perform relativistic density-functional calculations (DFT) with  FPLO v22.01-63 \cite{Koepernik1999} in the generalized gradient approximation \cite{Perdew1997}.
For Brillouin zone integrations we use a tetrahedron method  with a $k$-mesh having $19\times19\times17$ subdivisions.
To analyze the surface electronic structure, we study finite or semi-infinite slabs based on a Wannier Hamiltonian which includes Bi $6p$ and $6s$ orbitals together with Pt $6s$ and Pt $5d$ orbitals \cite{koepernik23}.

\begin{figure}[t!]
    \centering
    \includegraphics[width=\linewidth]{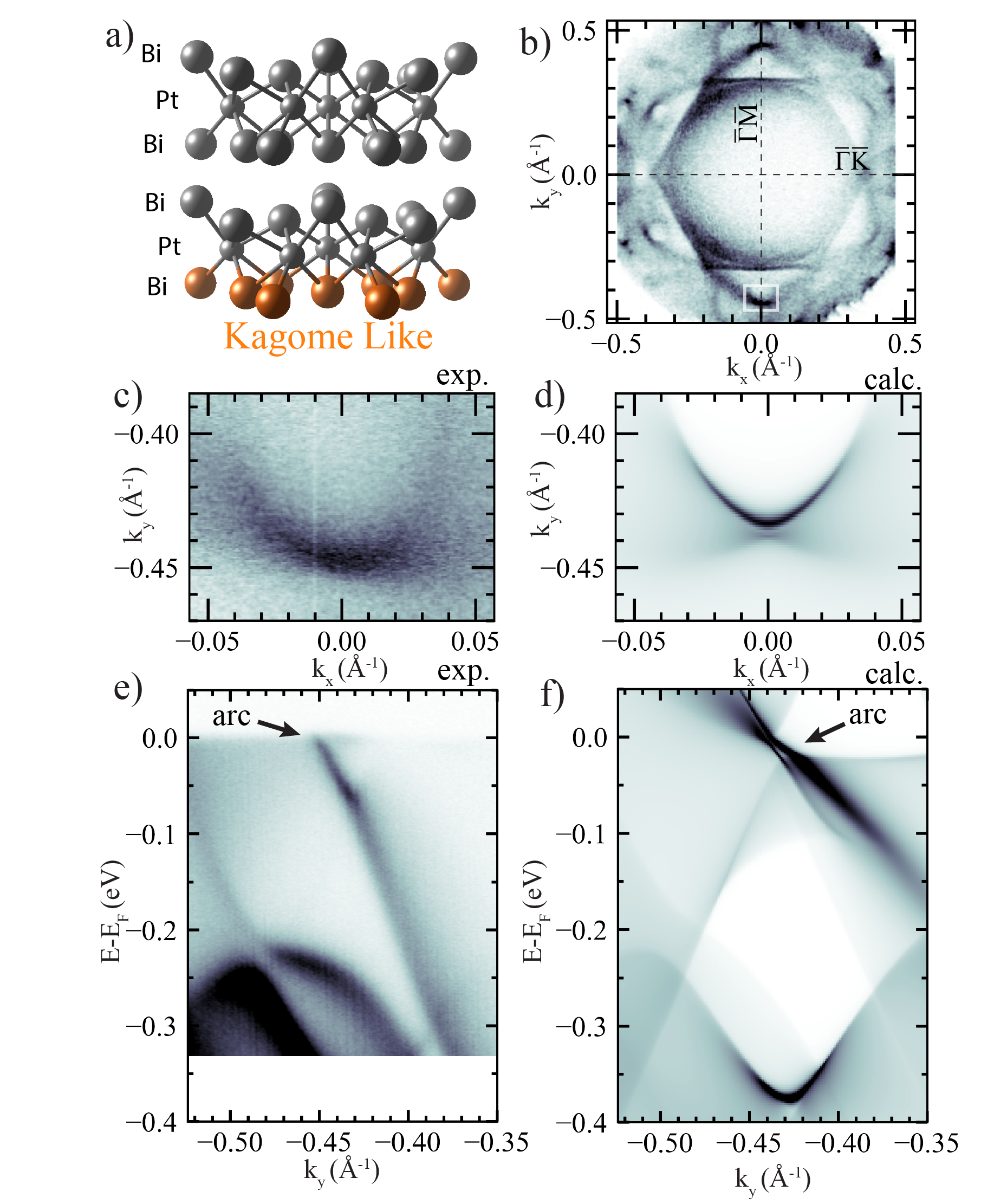}
    \caption{Same as Fig. 1 for the KL surface termination and with a photon energy of $h\nu$ = 18 eV for the experimental data.
    }
    \label{figure2}
\end{figure}

\textit{Low-energy fermiology--} The measured overview Fermi surface for the DH termination shows Fermi arcs along the six $\bar{\Gamma}\bar{M}$ lines of the surface Brillouin zone [Fig.~1(b)], in agreement with previous findings \cite{kuibarov_evidence_2024,oleary_topography_2025,vocaturo_electronic_2024,hoffmann2025fermi}. The Fermi arcs have a teardrop-like shape, as seen more clearly in Fig.~\ref{figure1}(c) and in a calculation of the surface spectral weight in Fig.~\ref{figure1}(d). Cuts through the Fermi-arc dispersion along $k_y$ ($\bar{\Gamma}\bar{M}$) from experiment and calculation are shown in Figs.~\ref{figure1}(e)-(f). The measured momentum shape, position, and dispersion of the arc match remarkably well with the calculation. Note that experimental and theoretical results are plotted on identical axis ranges. The tip of the arc has a Fermi wave-vector $k_\text{F} \approx 0.39$ Å$^{-1}$, in significant quantitative deviation from previous experimental results \cite{kuibarov_evidence_2024,kuibarov_measuring_2025}. Below the Fermi level the dispersion of the Fermi arc is shallow over its full momentum range, see also Fig.~3. 
The measured spectral weight of the arc reduces towards smaller $k_y$, which we attribute to increased mixing with bulk electronic states \cite{Min:2019}.

\begin{figure}[t!]
    \centering
    \includegraphics[width=1\linewidth]{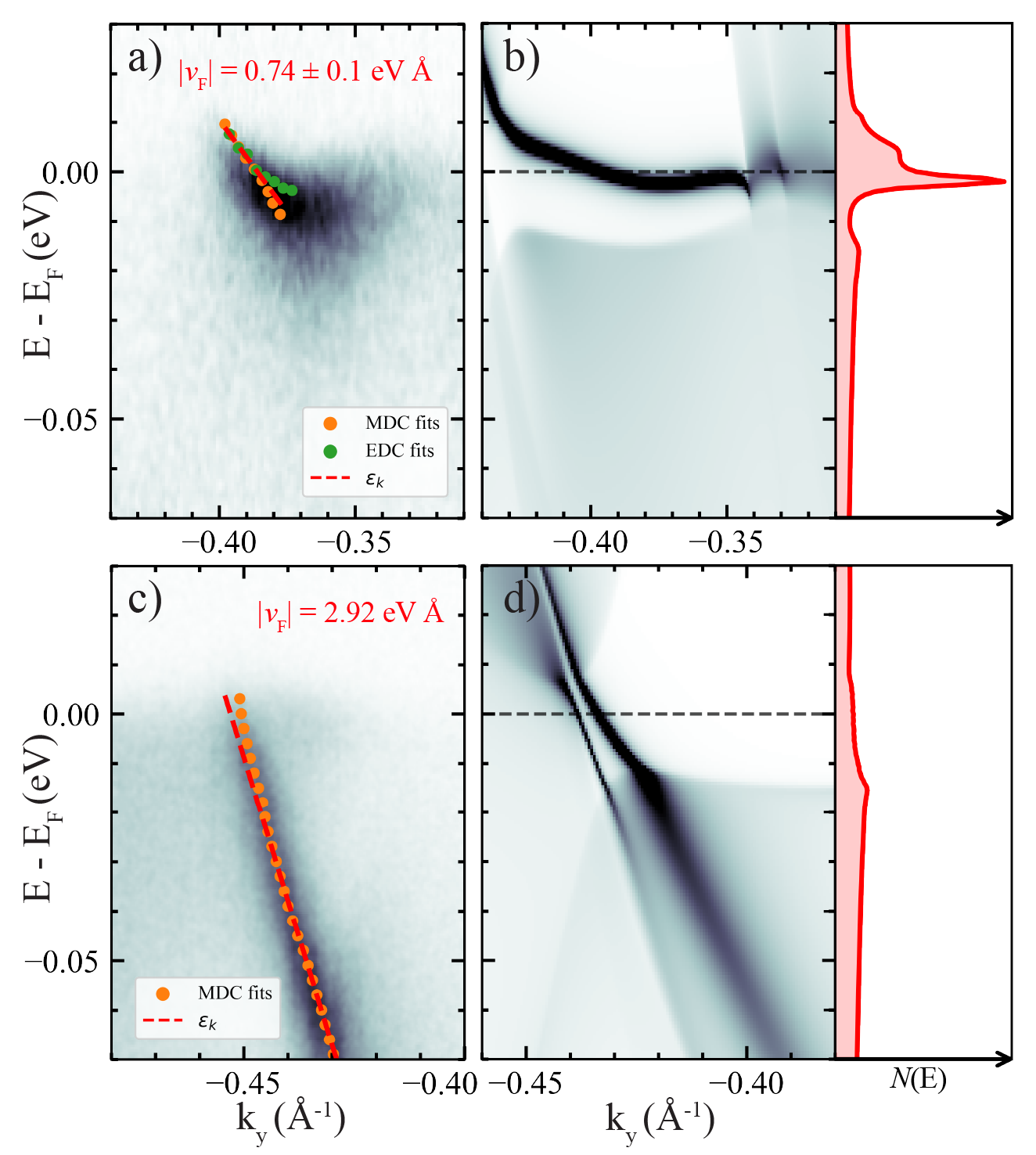}
    \caption{High resolution ARPES data along $\Gamma$-M through the center of the Fermi arc for a) the DH termination and c) the KL termination. The orange dots mark the peak positions determined from MDC cuts. In a) peak positions determined based on EDC cuts are also included in green. The red dashed line is a linear fit of the arc dispersion with the indicated Fermi velocity. Calculations of surface spectral weight for the same cuts as a) and c) are presented in b) for the DH termination and d) the KL termination. The Fermi-arc density of states is indicated by the red graphs in b), and d) showing that the DH termination has a higher density of states due to the flatter dispersion. }
    \label{figure3}
\end{figure}

Next we consider results for the KL termination [Fig.~\ref{figure2}(a)].
The Fermi surface again shows 6 Fermi arcs along the $\bar{\Gamma}\bar{M}$ directions [Fig.~\ref{figure2}(b)]. We compare measured and calculated momentum distributions of the Fermi arc in Figs.~\ref{figure2}(c)-(f). Different from the DH termination, the Fermi arc for the KL termination has a horseshoe shape. The tip of the arc is located at a significantly larger Fermi wave vector $k_\text{F} \approx 0.45$ $ \mbox{Å}^{-1}$, in good agreement between experiment and theory but, again, in substantial deviation to a previous report \cite{changdar_topological_2025}. The Fermi arc shows pronounced spectral blurring and reduction of spectral weight in momentum regions away from the tip of the arc. Our calculations indicate that this effect is related to the increased $k_\text{F}$ which implies a closer proximity to bulk states.

The most striking deviation from the DH termination is observed for the Fermi-arc dispersion [Fig. 2(e)-(f)]. We observe a steep, approximately linear dispersion of the Fermi arc, in contrast to the shallow dispersion for the DH termination [Fig. 1(e)-(f)]. As a result, the Fermi-arc bandwidth is much larger for the KL termination, extending over several 100 meV in binding energy. At approximately $E-E_{\text{F}}$ = -70 meV, we observe a kink-like change in slope of the Fermi-arc dispersion and increased spectral broadening, which are typical signatures of mixing with bulk states \cite{Bergman:13,Seibel:15}. This effect is nicely reproduced in our calculations in Fig.~2(f), albeit at lower binding energy. The steep dispersion and broad bandwidth of the KL termination has not been reported in previous ARPES studies \cite{changdar_topological_2025, kuibarov_evidence_2024, kuibarov_three_2025}. 

Figure \ref{figure3} compares the low-energy dispersion of the Fermi arcs for the two terminations. For the DH termination, the Fermi arc extends over only about 10 meV below the Fermi level [Fig.~\ref{figure3}(a)]. Based on fits to momentum and energy distribution curves (MDC and EDC), we estimate the Fermi velocity, $v_\text{F}$, to about 0.74 eVÅ, while the dispersion further flattens below the Fermi level. Comparing with the calculation in Fig.~\ref{figure3}(b), we find a nearly quantitative agreement in dispersion and location in momentum space. The flat dispersion of the Fermi arc gives rise to a sharp peak at the Fermi level in the density of states (DOS), as seen in Fig.~\ref{figure3}(b). This enhanced DOS at the Fermi level is beneficial for the formation of Cooper pairs in the superconducting state. For the KL termination, the Fermi-arc dispersion is significantly steeper [Fig.~\ref{figure3}(c)-(d)]. Based on MDC fits, we estimate the Fermi velocity to 2.92 eVÅ, nearly 4 times higher than on the DH termination. The DOS related to the Fermi arc is much lower than for the DH termination and shows no enhancement near the Fermi level [Fig.~\ref{figure3}(d)]. Note that in the calculation another feature is present to the left of the Fermi arc, which we attribute to a surface resonance. This feature is not discernible in the experimental data in Fig. 3(c), but a detailed analysis of momentum distributions supports its presence off the high-symmetry line [see Supplemental material \cite{suppl} Fig. S4]. 

The driving force for the different Fermi-arc dispersions is a termination-dependent coupling to bulk states. For the KL termination, the Fermi arc merges into the projected bulk band, located ca. 10-15 meV below the Fermi level, and maintains a steep dispersion [Figs.~\ref{figure3}(b) and (d)]. By contrast, for the DH termination, the Fermi-arc dispersion bends within the projected gap and evolves parallel to the weakly dispersive bulk continuum edge, resulting in a flat dispersion just below the Fermi level. 

We attribute the differences in surface-bulk coupling to differences in the atomic structure. The DH terminating atomic layer has higher in-plane symmetry ($C_{6v}$ for DH vs. $C_{3v}$ for KL), including a mirror plane along $\bar{\Gamma}\bar{M}$, which is absent in KL. Furthermore, the DH surface termination is buckled while the KL termination is planar. Both of these features influence orbital composition, orbital symmetry and depth dependence of states near the surface \cite{Chadi:78,Bertel:94,bentmann2009,Yan:15}. Interestingly, our calculations show a suppression of bulk-state weight near the DH surface but not near the KL surface, leading to weaker mixing with the surface states. We also find a complex momentum-dependent orbital composition for both terminations. At the Fermi level, the DH Fermi arc is dominated by in-plane Bi $p_x$ and $p_y$ orbitals, whereas the KL Fermi arc acquires a substantially enhanced $p_z$ character, further influencing the coupling to bulk states [see Supplemental Material \cite{suppl} Figs. S1, S2 and S3 for more details].

\begin{figure*}
    \centering
    \includegraphics[width=0.85\linewidth]{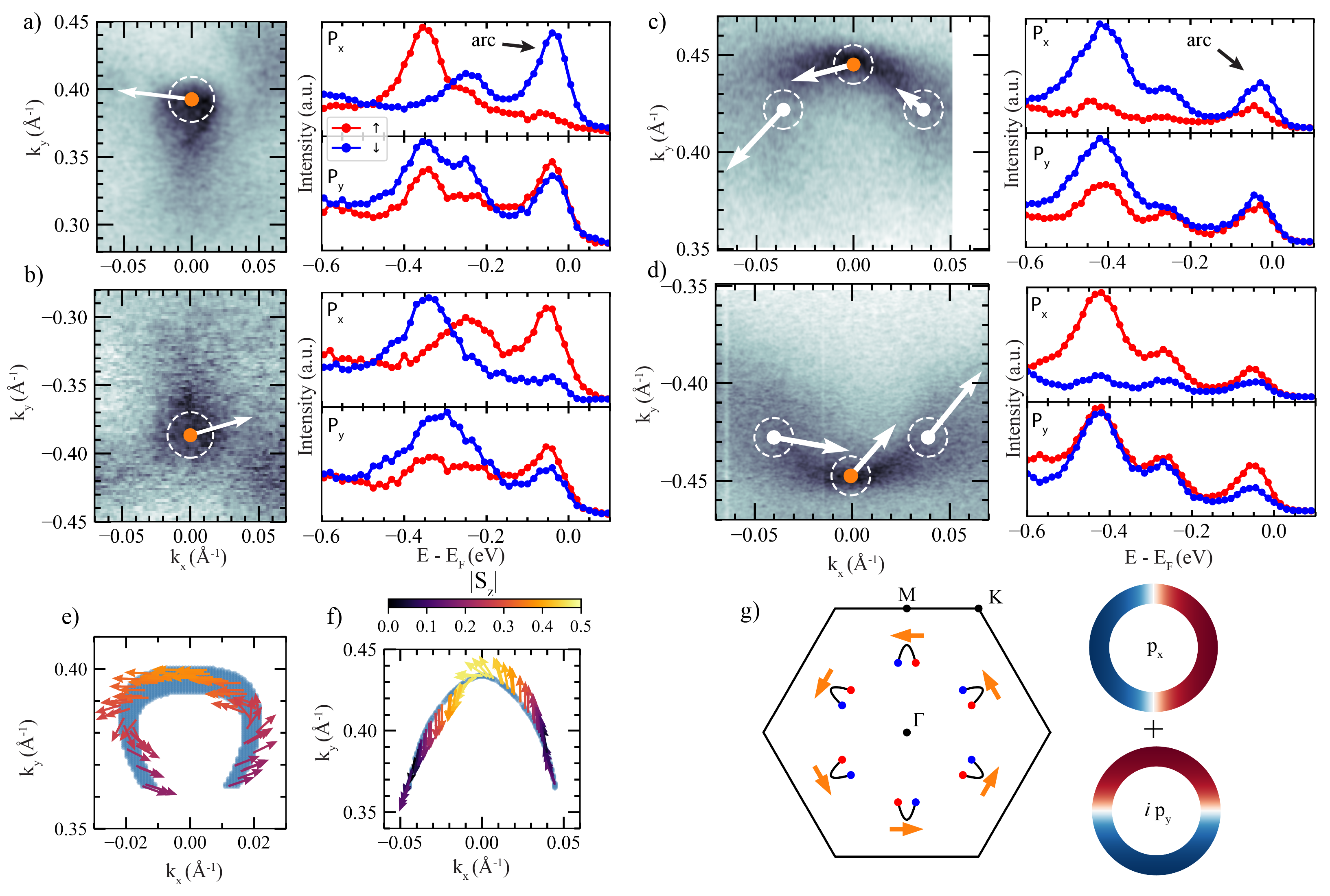}
    \caption{(a) Constant energy cut at the Fermi level for the DH termination with spin resolved energy distribution curves. The orange dot marks the position for the spin-resolved EDC. The dashed white circle indicates the momentum resolution for the spin measurements. The white arrow indicates the spin polarization. (b) Same measurement as in a) but for the Fermi arc on opposite side of $\Gamma$. c) and d) show the same as in a) and b) for the KL termination. For the DH (KL) termination a photon energy of 26 eV (18 eV) was used. e) and f) show calculations of the x-y spin texture of the Fermi arcs for the DH and KL termination, respectively. The arrows indicate the in-plane spin polarization direction, while the color encodes the out-of-plane component. g) Schematic of the spin texture of the Fermi arcs in PtBi$_2$. }
    \label{figure4}
\end{figure*}

\textit{Fermi-arc spin polarization--} We now move on to study the spin texture of the Fermi arcs. Figure \ref{figure4} shows spin-resolved measurements of the Fermi arcs at positive and negative $k_y$. Data for the DH termination is shown in Figs.~\ref{figure4}(a) and (b) and data for the KL termination is shown in Figs.~\ref{figure4}(c) and (d). The left panels show constant-energy maps of the Fermi arcs, measured in the spin-resolving setup, along with orange circles that mark the position where we captured spin-resolved EDC, shown in the right panel. The dashed white circle indicates the momentum resolution of the spin measurement. The right panels show spin-resolved EDC for spin quantization axes along $x$ (top) and along $y$ (bottom), giving the measured $P_x$ and $P_y$ components of the spin polarization vector, $\mathbf{P}$. From the Fermi level, the first peak in the EDC is the Fermi arc. The experimental data demonstrate a sizable spin polarization of the Fermi arcs on both terminations, showing that they form a single, non-degenerate state. A vectorial representation of the measured $P_x$ and $P_y$ Fermi-arc components is shown by the white arrows in the left panels. The peaks at higher binding energy correspond to other surface states [cf. Figs. 1 and 2] and also show significant spin polarization.  

For the DH termination in Figs.~\ref{figure4}(a)-(b), the measured Fermi-arc spin points mainly along the $x$ direction and is approximately opposite at $+k_y$ and $-k_y$. For the KL termination, we succeeded in probing multiple points along the Fermi arc, see Figs.~\ref{figure4}(c) and (d). The spin-resolved EDC corresponding to the white dots are given in Fig. S5 of the Supplemental Material \cite{suppl}. From the extracted arrows, we see that the spin is momentum-locked, as for the DH termination, and tends to curve along the arc. We note that for both terminations the spin forms an overall counterclockwise texture, as schematically depicted in Fig.~\ref{figure4}(g). 

The main observations in the spin-resolved measurements are supported by our calculations in Figs.~\ref{figure4}(e) and (f). In line with experiment, we find single-degeneracy and an overall helical spin-momentum locking of the Fermi arcs. Furthermore, our calculations indicate additional fine structure beyond the resolution of our experimental data. According to our calculations, both terminations show significant variation of the spin polarization with energy, including a finite out-of-plane component $P_z$, as well as a complex atomic-layer dependence with distance beneath the surface, similar to previous findings for topological insulators \cite{Zhu:13}. We attribute these effects to strong spin-orbit coupling in the Bi 6$p$ orbitals in combination with the complex surface-bulk hybridization discussed above.    

\textit{Discussion--} 
Our observations of spin polarization and termination-dependent low-energy dispersion of the Fermi arcs have implications for the properties of the superconducting state at the surface of PtBi$_2$ \cite{kuibarov_evidence_2024,huang_sizable_2025}. 
The singly degenerate nature of the Fermi arcs provides direct evidence for nontrivial topology. In particular, the measurements demonstrate that a single surface mode terminates at each projected Weyl node, which confirms the bulk-boundary-correspondence imposed by bulk Weyl nodes carrying a topological charge of $\pm 1$ \cite{Armitage:18}.
Furthermore, the momentum dependence of the spin wave function of the Bloch states on the Fermi arcs has important consequences for intrinsic topological superconductivity \cite{Maeland_2025_Phonon_mediated_top_supcon,maeland2025mechanismnodaltopologicalsuperconductivity}. We measure an in-plane spin winding that gives a momentum dependence of the Bloch states that can be classified as chiral $p$-wave [see Fig.~4(g)]. Recent theory predicts that the chiral $p$-wave momentum dependence of the Bloch states determines the momentum dependence of the superconducting gap ~\cite{Maeland_2025_Phonon_mediated_top_supcon}. In particular, for phonon-mediated pairing, a superconducting state of the spin-polarized Fermi arcs provides an intrinsic realization of the Fu–Kane model for topological superconductivity \cite{FuKane:07} realizing time-reversal symmetric spinless chiral $p$-wave superconductivity \cite{Scheurer:16}. Thus, together with previously reported superconductivity \cite{kuibarov_evidence_2024,huang_sizable_2025}, our observation of non-degeneracy and helical spin texture of the Fermi arcs provides support for the proposed intrinsic topological superconductivity in PtBi$_2$.

Theory predicts that also the surface-state bandwidth is an important ingredient in determining the nature of the superconducting state in PtBi$_2$ and related surface superconductors \cite{maeland2025mechanismnodaltopologicalsuperconductivity}. Our results demonstrate that the Fermi-arc bandwidth strongly depends on the surface termination. On the DH surface, we find a nearly flat Fermi-arc dispersion just below the Fermi level, giving rise to an enhanced surface DOS that is generally expected to strengthen superconductivity \cite{Maeland_2025_Phonon_mediated_top_supcon, Kopnin2011flatSC, Peotta2015flatSC, Maeland2023DecCC}.
In contrast, the KL surface exhibits a linear Fermi-arc dispersion without enhanced DOS. These results are in line with the prediction of a 50~\% larger superconducting-gap size for the DH compared to KL termination \cite{changdar_topological_2025}.

The termination dependence of the bandwidth may also influence the pairing symmetry. If the surface-state bandwidth is much larger than the maximum phonon energy, Morel Anderson physics \cite{Morel1962Anderson} implies that Coulomb repulsion only provides quantitative changes to phonon-mediated superconductivity. In contrast, qualitative changes in the gap symmetry require a small surface-state bandwidth comparable to the maximum phonon energy and were predicted \cite{maeland2025mechanismnodaltopologicalsuperconductivity} to underlie the observed $i$-wave superconductivity in PtBi$_2$ \cite{changdar_topological_2025}. Our results place the DH surface in the latter regime, whereas the KL surface appears to remain in the former. The observed termination dependence of the Fermi-arc bandwidth therefore suggests different pairing tendencies within a linearized gap-equation framework. However, the superconducting state ultimately realized may also be constrained globally by topology such that either none or both of the surfaces show $i$-wave superconductivity, as reported in Ref.~\cite{changdar_topological_2025}. Our results point to the possibility of a nontrivial interplay between local termination-dependent pairing tendencies and global topological constraints, whose consequences remain poorly understood.

\newpage

\textit{Conclusion--} We combined spin-ARPES and first-principles calculations to resolve the normal-state surface electronic structure of trigonal PtBi$_2$. We find a strong dependence on surface termination for the shape, position and dispersion of the Fermi arc surface states. The DH termination hosts a flat Fermi-arc dispersion, resulting in enhanced DOS near the Fermi energy. In contrast, the KL termination exhibits a steep Fermi-arc dispersion with much larger bandwidth. This establishes a unified picture of both surface terminations, which has so far been lacking. 
Based on first-principles calculations we attribute the contrasting dispersions to termination-dependent surface atomic structure and hybridization with bulk states. Our spin-resolved measurements demonstrate sizable spin polarization and non-degenerate Fermi arcs with helical spin-momentum locking. This establishes the non-trivial topology of the Fermi arcs and constitutes a key prerequisite for topological surface superconductivity in PtBi$_2$. 

\begin{acknowledgements}
\section{Acknowledgments}
This work was supported by the Research Council of Norway through Grant No. 323766 and its Centres of Excellence funding scheme Grant No. 262633 “QuSpin.” We acknowledge the MAX IV Laboratory for beamtime on the Bloch beamline under proposal No.20250217. Research conducted at MAX IV, a Swedish national user facility, is supported by Vetenskapsrådet (Swedish Research Council, VR) under contract 2018-07152, Vinnova (Swedish Governmental Agency for Innovation Systems) under contract 2018-04969 and Formas under contract 2019-02496. K.M.~and B.T.~were supported by the Deutsche Forschungsgemeinschaft (DFG, German Research Foundation) through SFB 1170 (project ID 258499086), and DFG through the W{\"u}rzburg-Dresden Cluster of Excellence ctd.qmat (EXC 2147, project ID 390858490). F.G.~was supported by the DFG through project ID 556350547. J.I.F. thanks Ulrike Nitzsche for technical assistance.

\end{acknowledgements}

\bibliography{ref}

@article{Scheurer:16,
  title = {Mechanism, time-reversal symmetry, and topology of superconductivity in noncentrosymmetric systems},
  author = {Scheurer, M. S.},
  journal = {Phys. Rev. B},
  volume = {93},
  issue = {17},
  pages = {174509},
  numpages = {25},
  year = {2016},
  month = {May},
  publisher = {American Physical Society},
  doi = {10.1103/PhysRevB.93.174509},
  url = {https://link.aps.org/doi/10.1103/PhysRevB.93.174509}
}

@article{Seibel:15,
  title = {Connection of a Topological Surface State with the Bulk Continuum in {Sb$_2$Te$_3$(0001)}},
  author = {Seibel, Christoph and Bentmann, Hendrik and Braun, J\"urgen and Min\'ar, Jan and Maa\ss{}, Henriette and Sakamoto, Kazuyuki and Arita, Masashi and Shimada, Kenya and Ebert, Hubert and Reinert, Friedrich},
  journal = {Phys. Rev. Lett.},
  volume = {114},
  issue = {6},
  pages = {066802},
  numpages = {5},
  year = {2015}
}

@article{Armitage:18,
  title = {Weyl and Dirac semimetals in three-dimensional solids},
  author = {Armitage, N. P. and Mele, E. J. and Vishwanath, Ashvin},
  journal = {Rev. Mod. Phys.},
  volume = {90},
  issue = {1},
  pages = {015001},
  numpages = {57},
  year = {2018},
  month = {Jan},
  publisher = {American Physical Society},
  doi = {10.1103/RevModPhys.90.015001},
  url = {https://link.aps.org/doi/10.1103/RevModPhys.90.015001}
}

@article{Yan:15,
  title = {Topological surface states and {Fermi} arcs of the noncentrosymmetric {Weyl} semimetals {TaAs}, {TaP}, {NbAs}, and {NbP}},
  author = {Sun, Yan and Wu, Shu-Chun and Yan, Binghai},
  journal = {Phys. Rev. B},
  volume = {92},
  issue = {11},
  pages = {115428},
  numpages = {11},
  year = {2015},
  month = {Sep},
  publisher = {American Physical Society},
  doi = {10.1103/PhysRevB.92.115428},
  url = {https://link.aps.org/doi/10.1103/PhysRevB.92.115428}
}

@article{Bertel:94,
  title = {Symmetry of surface states},
  author = {Bertel, E.},
  journal = {Phys. Rev. B},
  volume = {50},
  issue = {7},
  pages = {4925--4928},
  numpages = {0},
  year = {1994},
  month = {Aug},
  publisher = {American Physical Society},
  doi = {10.1103/PhysRevB.50.4925},
  url = {https://link.aps.org/doi/10.1103/PhysRevB.50.4925}
}

@article{Chadi:78,
  title = {(110) surface states of {GaAs}: Sensitivity of electronic structure to surface structure},
  author = {Chadi, D. J.},
  journal = {Phys. Rev. B},
  volume = {18},
  issue = {4},
  pages = {1800--1812},
  numpages = {0},
  year = {1978},
  month = {Aug},
  publisher = {American Physical Society},
  doi = {10.1103/PhysRevB.18.1800},
  url = {https://link.aps.org/doi/10.1103/PhysRevB.18.1800}
}

@article{bentmann2009,
  title={Origin and manipulation of the Rashba splitting in surface alloys},
  author={Bentmann, H and Forster, F and Bihlmayer, G and Chulkov, EV and Moreschini, L and Grioni, M and Reinert, F},
  journal={EPL},
  volume={87},
  number={3},
  pages={37003},
  year={2009}
}

@article{Bergman:13,
  title = {Bulk metals with helical surface states},
  author = {Bergman, Doron L. and Refael, Gil},
  journal = {Phys. Rev. B},
  volume = {82},
  issue = {19},
  pages = {195417},
  numpages = {11},
  year = {2010}
}

@article{Zhu:13,
  title = {Layer-By-Layer Entangled Spin-Orbital Texture of the Topological Surface State in {Bi$_2$Se$_3$}},
  author = {Zhu, Z.-H. and Veenstra, C. N. and Levy, G. and Ubaldini, A. and Syers, P. and Butch, N. P. and Paglione, J. and Haverkort, M. W. and Elfimov, I. S. and Damascelli, A.},
  journal = {Phys. Rev. Lett.},
  volume = {110},
  issue = {21},
  pages = {216401},
  numpages = {5},
  year = {2013},
  month = {May},
  publisher = {American Physical Society},
  doi = {10.1103/PhysRevLett.110.216401},
  url = {https://link.aps.org/doi/10.1103/PhysRevLett.110.216401}
}

@article{FuKane:07,
  title = {Superconducting Proximity Effect and Majorana Fermions at the Surface of a Topological Insulator},
  author = {Fu, Liang and Kane, C. L.},
  journal = {Phys. Rev. Lett.},
  volume = {100},
  issue = {9},
  pages = {096407},
  numpages = {4},
  year = {2008},
  month = {Mar},
  publisher = {American Physical Society},
  doi = {10.1103/PhysRevLett.100.096407},
  url = {https://link.aps.org/doi/10.1103/PhysRevLett.100.096407}
}

@article{kuibarov_evidence_2024,
	title = {Evidence of superconducting {Fermi} arcs},
	volume = {626},
	issn = {1476-4687},
	url = {https://doi.org/10.1038/s41586-023-06977-7},
	doi = {10.1038/s41586-023-06977-7},
	number = {7998},
	journal = {Nature},
	author = {Kuibarov, Andrii and Suvorov, Oleksandr and Vocaturo, Riccardo and Fedorov, Alexander and Lou, Rui and Merkwitz, Luise and Voroshnin, Vladimir and Facio, Jorge I. and Koepernik, Klaus and Yaresko, Alexander and Shipunov, Grigory and Aswartham, Saicharan and Brink, Jeroen van den and Büchner, Bernd and Borisenko, Sergey},
	month = feb,
	year = {2024},
	pages = {294--299},
}

@article{changdar_topological_2025,
	title = {Topological nodal i-wave superconductivity in {PtBi$_2$}},
	volume = {647},
	copyright = {2025 The Author(s)},
	issn = {1476-4687},
	url = {https://www.nature.com/articles/s41586-025-09712-6},
	doi = {10.1038/s41586-025-09712-6},
	number = {8090},
	urldate = {2025-12-12},
	journal = {Nature},
	author = {Changdar, Susmita and Suvorov, Oleksandr and Kuibarov, Andrii and Thirupathaiah, Setti and Shipunov, Grigory and Aswartham, Saicharan and Wurmehl, Sabine and Kovalchuk, Iryna and Koepernik, Klaus and Timm, Carsten and Büchner, Bernd and Fulga, Ion Cosma and Borisenko, Sergey and Brink, Jeroen van den},
	month = nov,
	year = {2025},
	pages = {613--618},
}

@article{schimmel_surface_2024,
	title = {Surface superconductivity in the topological {Weyl} semimetal t-{PtBi$_2$}},
	volume = {15},
	copyright = {2024 The Author(s)},
	issn = {2041-1723},
	url = {https://www.nature.com/articles/s41467-024-54389-6},
	doi = {10.1038/s41467-024-54389-6},
	number = {1},
	urldate = {2025-12-12},
	journal = {Nat. Commun.},
	author = {Schimmel, Sebastian and Fasano, Yanina and Hoffmann, Sven and Besproswanny, Julia and Corredor Bohorquez, Laura Teresa and Puig, Joaquín and Elshalem, Bat-Chen and Kalisky, Beena and Shipunov, Grigory and Baumann, Danny and Aswartham, Saicharan and Büchner, Bernd and Hess, Christian},
	month = nov,
	year = {2024},
	pages = {9895},

}

@article{mandal2023topological,
  title={Topological superconductors from a materials perspective},
  author={Mandal, Manasi and Drucker, Nathan C and Siriviboon, Phum and Nguyen, Thanh and Boonkird, Artittaya and Lamichhane, Tej Nath and Okabe, Ryotaro and Chotrattanapituk, Abhijatmedhi and Li, Mingda},
  journal={Chem. Mater.},
  volume={35},
  number={16},
  pages={6184--6200},
  year={2023},
  publisher={ACS Publications}
}

@article{Min:2019,
  title = {Orbital Fingerprint of Topological {Fermi} Arcs in the {Weyl} Semimetal {TaP}},
  author = {Min, Chul-Hee and Bentmann, Hendrik and Neu, Jennifer N. and Eck, Philipp and Moser, Simon and Figgemeier, Tim and \"Unzelmann, Maximilian and Kissner, Katharina and Lutz, Peter and Koch, Roland J. and Jozwiak, Chris and Bostwick, Aaron and Rotenberg, Eli and Thomale, Ronny and Sangiovanni, Giorgio and Siegrist, Theo and Di Sante, Domenico and Reinert, Friedrich},
  journal = {Phys. Rev. Lett.},
  volume = {122},
  issue = {11},
  pages = {116402},
  numpages = {6},
  year = {2019},
  month = {Mar},
  publisher = {American Physical Society},
  doi = {10.1103/PhysRevLett.122.116402},
  url = {https://link.aps.org/doi/10.1103/PhysRevLett.122.116402}
}

@article{hoffmann2025fermi,
  title={Fermi arcs dominating the electronic surface properties of trigonal {PtBi$_2$}},
  author={Hoffmann, Sven and Schimmel, Sebastian and Vocaturo, Riccardo and Puig, Joaquin and Shipunov, Grigory and Janson, Oleg and Aswartham, Saicharan and Baumann, Danny and B{\"u}chner, Bernd and Brink, Jeroen van den and others},
  journal={Adv. Phys. Res.},
  volume={4},
  number={5},
  pages={2400150},
  year={2025},
  publisher={Wiley Online Library}
}

@article{RevModPhys.80.1083,
  title = {Non-Abelian anyons and topological quantum computation},
  author = {Nayak, Chetan and Simon, Steven H. and Stern, Ady and Freedman, Michael and Das Sarma, Sankar},
  journal = {Rev. Mod. Phys.},
  volume = {80},
  issue = {3},
  pages = {1083--1159},
  numpages = {0},
  year = {2008},
  month = {Sep},
  publisher = {American Physical Society},
  doi = {10.1103/RevModPhys.80.1083},
  url = {https://link.aps.org/doi/10.1103/RevModPhys.80.1083}
}

@article{Lee:24,
  title = {Fermi Surface Spin Texture and Topological Superconductivity in Spin-Orbit Free Noncollinear Antiferromagnets},
  author = {Lee, Seung Hun and Qian, Yuting and Yang, Bohm-Jung},
  journal = {Phys. Rev. Lett.},
  volume = {132},
  issue = {19},
  pages = {196602},
  numpages = {6},
  year = {2024},
  month = {May},
  publisher = {American Physical Society},
  doi = {10.1103/PhysRevLett.132.196602},
  url = {https://link.aps.org/doi/10.1103/PhysRevLett.132.196602}
}

@article{KobayashiSato2015,
  author       = {Shingo Kobayashi and Masatoshi Sato},
  title        = {Topological superconductivity in Dirac semimetals},
  journal      = {Phys. Rev. Lett.},
  volume       = {115},
  pages        = {187001},
  year         = {2015},
  doi          = {10.1103/PhysRevLett.115.187001},
  url          = {https://doi.org/10.1103/PhysRevLett.115.187001}
}

@article{SatoAndo2017,
  author       = {Masatoshi Sato and Yoichi Ando},
  title        = {Topological superconductors: a review},
  journal      = {Rep. Prog. Phys.},
  volume       = {80},
  number       = {7},
  pages        = {076501},
  year         = {2017},
  doi          = {10.1088/1361-6633/aa6ac7},
  url          = {https://doi.org/10.1088/1361-6633/aa6ac7}
}

@article{frolov2020topological,
  title={Topological superconductivity in hybrid devices},
  author={Frolov, SM and Manfra, MJ and Sau, JD},
  journal={Nat. Phys.},
  volume={16},
  number={7},
  pages={718--724},
  year={2020},
  publisher={Nature Publishing Group UK London}
}

@article{keimer2017physics,
  title={The physics of quantum materials},
  author={Keimer, Bernhard and Moore, Joel E},
  journal={Nat. Phys.},
  volume={13},
  number={11},
  pages={1045--1055},
  year={2017},
  publisher={Nature Publishing Group UK London}
}

@article{di2025kagome,
  title={Kagome metals},
  author={Di Sante, Domenico and Neupert, Titus and Sangiovanni, Giorgio and Thomale, Ronny and Comin, Riccardo and Zeljkovic, Ilija and Checkelsky, Joseph G and Wilson, Stephen D},
  journal={arXiv preprint arXiv:2511.12731},
  year={2025}
}

@article{bernevig2022progress,
  title={Progress and prospects in magnetic topological materials},
  author={Bernevig, B Andrei and Felser, Claudia and Beidenkopf, Haim},
  journal={Nature},
  volume={603},
  number={7899},
  pages={41--51},
  year={2022},
  publisher={Nature Publishing Group UK London}
}

@article{Shipunov_polymorphic_PtBi2_2020,
  title = {Polymorphic ${\mathrm{PtBi}}_{2}$: Growth, structure, and superconducting properties},
  author = {Shipunov, G. and Kovalchuk, I. and Piening, B. R. and Labracherie, V. and Veyrat, A. and Wolf, D. and Lubk, A. and Subakti, S. and Giraud, R. and Dufouleur, J. and Shokri, S. and Caglieris, F. and Hess, C. and Efremov, D. V. and B\"uchner, B. and Aswartham, S.},
  journal = {Phys. Rev. Mater.},
  volume = {4},
  issue = {12},
  pages = {124202},
  numpages = {8},
  year = {2020},
  month = {Dec},
  publisher = {American Physical Society},
  doi = {10.1103/PhysRevMaterials.4.124202},
  url = {https://link.aps.org/doi/10.1103/PhysRevMaterials.4.124202}
}

@article{huang_sizable_2025,
  title={Sizable superconducting gap and anisotropic chiral topological superconductivity in the Weyl semimetal PtBi $ \_2$},
  author={Huang, Xiaochun and Zhao, Lingxiao and Schimmel, Sebastian and Besproswanny, Julia and H{\"a}rtl, Patrick and Hess, Christian and B{\"u}chner, Bernd and Bode, Matthias},
  journal={arXiv preprint arXiv:2507.13843},
  year={2025}
}

@article{oleary_topography_2025,
	title = {Topography of {Fermi} arcs in {$t$-PtBi$_2$} using high-resolution angle-resolved photoemission spectroscopy},
	volume = {112},
	url = {https://link.aps.org/doi/10.1103/n5pz-j2sl},
	doi = {10.1103/n5pz-j2sl},
	number = {8},
	urldate = {2025-12-12},
	journal = {Phys. Rev. B},
	author = {O'Leary, Evan and Li, Zhuoqi and Wang, Lin-Lin and Schrunk, Benjamin and Eaton, Andrew and Canfield, Paul C. and Kaminski, Adam},
	month = aug,
	year = {2025},
	pages = {085154},
}

@misc{kuibarov_three_2025,
	title = {Three prerequisites for high-temperature superconductivity in t-{PtBi$_2$}},
	url = {http://arxiv.org/abs/2509.02178},
	doi = {10.48550/arXiv.2509.02178},
	urldate = {2025-12-16},
	publisher = {arXiv},
	author = {Kuibarov, Andrii and Changdar, Susmita and Vocaturo, Riccardo and Suvorov, Oleksandr and Fedorov, Alexander and Lou, Rui and Krivenkov, Maxim and Harnagea, Luminita and Wurmehl, Sabine and Brink, Jeroen van den and Büchner, Bernd and Borisenko, Sergey},
	month = sep,
	year = {2025},
	note = {arXiv:2509.02178 [cond-mat] version: 1},
}

@article{kuibarov_measuring_2025,
	title = {Measuring superconducting arcs by angle-resolved photoemission spectroscopy},
	volume = {112},
	url = {https://link.aps.org/doi/10.1103/5q36-wgl9},
	doi = {10.1103/5q36-wgl9},
	number = {14},
	urldate = {2025-12-12},
	journal = {Phys. Rev. B},
	author = {Kuibarov, A. and Changdar, S. and Fedorov, A. and Lou, R. and Suvorov, O. and Misheneva, V. and Harnagea, L. and Kovalchuk, I. and Wurmehl, S. and Büchner, B. and Borisenko, S.},
	month = oct,
	year = {2025},
	pages = {144518},
}

@article{vocaturo_electronic_2024,
	title = {Electronic structure of the surface-superconducting {Weyl} semimetal {PtBi$_2$}},
	volume = {110},
	url = {https://link.aps.org/doi/10.1103/PhysRevB.110.054504},
	doi = {10.1103/PhysRevB.110.054504},
	number = {5},
	urldate = {2025-12-17},
	journal = {Phys. Rev. B},
	author = {Vocaturo, Riccardo and Koepernik, Klaus and Facio, Jorge I. and Timm, Carsten and Fulga, Ion Cosma and Janson, Oleg and van den Brink, Jeroen},
	month = aug,
	year = {2024},
	pages = {054504},
}

@article{Yang_Giant_linear_magneto_resistance_2016,
    author = {Yang, Xiaojun and Bai, Hua and Wang, Zhen and Li, Yupeng and Chen, Qian and Chen, Jian and Li, Yuke and Feng, Chunmu and Zheng, Yi and Xu, Zhu-an},
    title = {Giant linear magneto-resistance in nonmagnetic {PtBi$_2$}},
    journal = {Appl. Phys. Lett.},
    volume = {108},
    number = {25},
    pages = {252401},
    year = {2016},
    month = {06},
    issn = {0003-6951},
    doi = {10.1063/1.4954272},
    url = {https://doi.org/10.1063/1.4954272},
}

@article{BISWAS_1969_PtBi2,
title = {Strukturuntersuchungen in den mischungen {Pt-Tl-Pb} und {Pt-Pb-Bi}},
journal = {J. Less-Common Met.},
volume = {19},
number = {3},
pages = {223-243},
year = {1969},
issn = {0022-5088},
doi = {https://doi.org/10.1016/0022-5088(69)90099-X},
url = {https://www.sciencedirect.com/science/article/pii/002250886990099X},
author = {T. Biswas and K. Schubert}
}

@article{Yao_Bulk_and_Surface_2016,
  title = {Bulk and surface electronic structure of hexagonal structured {PtBi$_2$} studied by angle-resolved photoemission spectroscopy},
  author = {Yao, Q. and Du, Y. P. and Yang, X. J. and Zheng, Y. and Xu, D. F. and Niu, X. H. and Shen, X. P. and Yang, H. F. and Dudin, P. and Kim, T. K. and Hoesch, M. and Vobornik, I. and Xu, Z.-A. and Wan, X. G. and Feng, D. L. and Shen, D. W.},
  journal = {Phys. Rev. B},
  volume = {94},
  issue = {23},
  pages = {235140},
  numpages = {6},
  year = {2016},
  month = {Dec},
  publisher = {American Physical Society},
  doi = {10.1103/PhysRevB.94.235140},
  url = {https://link.aps.org/doi/10.1103/PhysRevB.94.235140}
}

@article{Kaiser_PtBi2_Low_T_reduction_2014,
author = {Kaiser, Martin and Baranov, Alexey I. and Ruck, Michael},
title = {{Bi$_2$Pt($hP9$)} by Low-Temperature Reduction of {Bi$_{13}$Pt$_3$I$_7$}: Reinvestigation of the Crystal Structure and Chemical Bonding Analysis},
journal = {Z. anorg. allg. Chem.},
volume = {640},
number = {14},
pages = {2742-2746},
doi = {https://doi.org/10.1002/zaac.201400331},
url = {https://onlinelibrary.wiley.com/doi/abs/10.1002/zaac.201400331},
year = {2014}
}

@article{gao_tripple_degenerate_2018,
	title = {A possible candidate for triply degenerate point fermions in trigonal layered {PtBi$_2$}},
	volume = {9},
	issn = {2041-1723},
	url = {https://doi.org/10.1038/s41467-018-05730-3},
	doi = {10.1038/s41467-018-05730-3},
	number = {1},
	journal = {Nat. Commun.},
	author = {Gao, Wenshuai and Zhu, Xiangde and Zheng, Fawei and Wu, Min and Zhang, Jinglei and Xi, Chuanying and Zhang, Ping and Zhang, Yuheng and Hao, Ning and Ning, Wei and Tian, Mingliang},
	month = aug,
	year = {2018},
	pages = {3249},
}

@article{Thirupathaiah_Origin_of_MR_2018,
  title = {Possible origin of linear magnetoresistance: Observation of {Dirac} surface states in layered {PtBi$_2$}},
  author = {Thirupathaiah, S. and Kushnirenko, Y. and Haubold, E. and Fedorov, A. V. and Rienks, E. D. L. and Kim, T. K. and Yaresko, A. N. and Blum, C. G. F. and Aswartham, S. and B\"uchner, B. and Borisenko, S. V.},
  journal = {Phys. Rev. B},
  volume = {97},
  issue = {3},
  pages = {035133},
  numpages = {6},
  year = {2018},
  month = {Jan},
  publisher = {American Physical Society},
  doi = {10.1103/PhysRevB.97.035133},
  url = {https://link.aps.org/doi/10.1103/PhysRevB.97.035133}
}

@article{feng_rashba-like_2019,
	title = {Rashba-like spin splitting along three momentum directions in trigonal layered {PtBi$_2$}},
	volume = {10},
	issn = {2041-1723},
	url = {https://doi.org/10.1038/s41467-019-12805-2},
	doi = {10.1038/s41467-019-12805-2},
	number = {1},
	journal = {Nat. Commun.},
	author = {Feng, Ya and Jiang, Qi and Feng, Baojie and Yang, Meng and Xu, Tao and Liu, Wenjing and Yang, Xiufu and Arita, Masashi and Schwier, Eike F. and Shimada, Kenya and Jeschke, Harald O. and Thomale, Ronny and Shi, Youguo and Wu, Xianxin and Xiao, Shaozhu and Qiao, Shan and He, Shaolong},
	month = oct,
	year = {2019},
	pages = {4765},
}

@article{veyrat_berezinskiikosterlitzthouless_2023,
	title = {Berezinskii–{Kosterlitz}–{Thouless} {Transition} in the {Type}-{I} {Weyl} {Semimetal} {PtBi$_2$}},
	volume = {23},
	issn = {1530-6984},
	url = {https://doi.org/10.1021/acs.nanolett.2c04297},
	doi = {10.1021/acs.nanolett.2c04297},
	number = {4},
	journal = {Nano Lett.},
	author = {Veyrat, Arthur and Labracherie, Valentin and Bashlakov, Dima L. and Caglieris, Federico and Facio, Jorge I. and Shipunov, Grigory and Charvin, Titouan and Acharya, Rohith and Naidyuk, Yurii and Giraud, Romain and van den Brink, Jeroen and Büchner, Bernd and Hess, Christian and Aswartham, Saicharan and Dufouleur, Joseph},
	month = feb,
	year = {2023},
	pages = {1229--1235},
}

@article{Bashlakov_2022_El_Ph_SupCon,
    author = {Bashlakov, D. L. and Kvitnitskaya, O. E. and Shipunov, G. and Aswartham, S. and Feya, O. D. and Efremov, D. V. and Büchner, B. and Naidyuk, Yu. G.},
    title = {Electron-phonon interaction and point contact enhanced superconductivity in trigonal {PtBi$_2$}},
    journal = {Low Temp. Phys.},
    volume = {48},
    number = {10},
    pages = {747-754},
    year = {2022},
    month = {10},
    issn = {1063-777X},
    doi = {10.1063/10.0014014},
    url = {https://doi.org/10.1063/10.0014014},
}

@article{Wang_2021_Press_Induced_SupCon,
  title = {Pressure-induced superconductivity in trigonal layered {PtBi$_2$} with triply degenerate point fermions},
  author = {Wang, Jing and Chen, Xuliang and Zhou, Yonghui and An, Chao and Zhou, Ying and Gu, Chuanchuan and Tian, Mingliang and Yang, Zhaorong},
  journal = {Phys. Rev. B},
  volume = {103},
  issue = {1},
  pages = {014507},
  numpages = {6},
  year = {2021},
  month = {Jan},
  publisher = {American Physical Society},
  doi = {10.1103/PhysRevB.103.014507},
  url = {https://link.aps.org/doi/10.1103/PhysRevB.103.014507}
}

@article{Wenxiang_2020_electronic_structure_of_PtBi2,
    author = {Jiang, Wenxiang and Zhu, Fengfeng and Li, Ping and Li, Yunlong and Wang, Guohua and Jing, Qiang and Gao, Wenshuai and Tian, Mingliang and Ma, Jie and Zhang, Wentao and Luo, Weidong and Qian, Dong},
    title = {Electronic structure of non-centrosymmetric {PtBi$_2$} studied by angle-resolved photoemission spectroscopy},
    journal = {J. Appl. Phys.},
    volume = {128},
    number = {13},
    pages = {135103},
    year = {2020},
    month = {10},
    issn = {0021-8979},
    doi = {10.1063/5.0020622},
    url = {https://doi.org/10.1063/5.0020622},
}

@article{Maeland_2025_Phonon_mediated_top_supcon,
  title = {Phonon-mediated intrinsic topological superconductivity in {Fermi} arcs},
  author = {M\ae{}land, Kristian and Bahari, Masoud and Trauzettel, Bj\"orn},
  journal = {Phys. Rev. B},
  volume = {112},
  issue = {10},
  pages = {104507},
  numpages = {15},
  year = {2025},
  month = {Sep},
  publisher = {American Physical Society},
  doi = {10.1103/47vs-qgzk},
  url = {https://link.aps.org/doi/10.1103/47vs-qgzk}
}

@misc{maeland2025mechanismnodaltopologicalsuperconductivity,
      title={Mechanism for Nodal Topological Superconductivity on {PtBi$_2$} Surface}, 
      author={Kristian Mæland and Giorgio Sangiovanni and Björn Trauzettel},
      year={2025},
      eprint={2512.09994},
      archivePrefix={arXiv},
      primaryClass={cond-mat.supr-con},
      url={https://arxiv.org/abs/2512.09994}, 
}

@article{Maeland2023DecCC,
	author = {M{\ae}land, Kristian and Sudb{\o}, Asle},
	title = {{Exceeding the Chandrasekhar-Clogston limit in flat-band superconductors: A multiband strong-coupling approach}},
	journal = {Phys. Rev. B},
	volume = {108},
	number = {21},
	pages = {214511},
	year = {2023},
	month = dec,
	issn = {2469-9969},
	publisher = {American Physical Society},
	url = {https://link.aps.org/doi/10.1103/PhysRevB.108.214511},
	doi = {10.1103/PhysRevB.108.214511}
}

@article{Peotta2015flatSC,
	author = {Peotta, Sebastiano and T{\ifmmode\ddot{o}\else\"{o}\fi}rm{\ifmmode\ddot{a}\else\"{a}\fi}, P{\ifmmode\ddot{a}\else\"{a}\fi}ivi},
	title = {{Superfluidity in topologically nontrivial flat bands}},
	journal = {Nat. Commun.},
	volume = {6},
	number = {8944},
	pages = {8944},
	year = {2015},
	month = nov,
	issn = {2041-1723},
	publisher = {Nature Publishing Group},
	doi = {10.1038/ncomms9944}
}

@article{Kopnin2011flatSC,
	author = {Kopnin, N. B. and Heikkil{\ifmmode\ddot{a}\else\"{a}\fi}, T. T. and Volovik, G. E.},
	title = {{High-temperature surface superconductivity in topological flat-band systems}},
	journal = {Phys. Rev. B},
	volume = {83},
	number = {22},
	pages = {220503},
	year = {2011},
	month = jun,
	publisher = {American Physical Society},
	doi = {10.1103/PhysRevB.83.220503}
}

@article{Dsouza2026KL,
	author = {Dsouza, Reuel and Parthenios, Nikolaos and Andersen, Brian M. and Christensen, Morten H.},
	title = {{Kohn-Luttinger Superconductivity of Weyl Fermi Arcs in PtBi$_2$}},
	journal = {arXiv:2605.31501},
	year = {2026},
	month = may,
	url = {https://doi.org/10.48550/arXiv.2605.31501}
}

@article{Buccheri2026ph,
	author = {Buccheri, Francesco and de Martino, Alessandro and Brink, Jeroen van den},
	title = {{Phonon-driven nodal surface superconductivity of Fermi arcs}},
	journal = {arXiv:2606.02371},
	year = {2026},
	month = jun,
	url = {https://doi.org/10.48550/arXiv.2606.02371}
}

@article{Morel1962Anderson,
	author = {Morel, P. and Anderson, P. W.},
	title = {{Calculation of the Superconducting State Parameters with Retarded Electron-Phonon Interaction}},
	journal = {Phys. Rev.},
	volume = {125},
	number = {4},
	pages = {1263--1271},
	year = {1962},
	month = feb,
	publisher = {American Physical Society},
	doi = {10.1103/PhysRev.125.1263}
}

@article{Koepernik1999,
  doi = {10.1103/physrevb.59.1743},
  url = {https://doi.org/10.1103/physrevb.59.1743},
  year = {1999},
  month = jan,
  publisher = {American Physical Society ({APS})},
  volume = {59},
  number = {3},
  pages = {1743--1757},
  author = {Klaus Koepernik and Helmut Eschrig},
  title = {Full-potential nonorthogonal local-orbital minimum-basis band-structure scheme},
  journal = {Phys. Rev. B},
}

@article{Perdew1997,
  title = {Generalized Gradient Approximation Made Simple},
  author = {Perdew, John P. and Burke, Kieron and Ernzerhof, Matthias},
  journal = {Phys. Rev. Lett.},
  volume = {78},
  pages = {1396--1396},
  year = {1997},
  doi = {10.1103/PhysRevLett.78.1396},
}

@article{koepernik23,
  title = {{Symmetry-conserving maximally projected {Wannier} functions}},
  author = {Koepernik, K. and Janson, O. and Sun, Yan and van den Brink, J.},
  journal = {Phys. Rev. B},
  volume = {107},
  pages = {235135},
  year = {2023},
  doi = {10.1103/PhysRevB.107.235135},
}

@article{Nomani2023FermiArcSCDOS,
	author = {Nomani, Aymen and Hosur, Pavan},
	title = {{Intrinsic surface superconducting instability in type-I Weyl semimetals}},
	journal = {Phys. Rev. B},
	volume = {108},
	number = {16},
	pages = {165144},
	year = {2023},
	month = oct,
	issn = {2469-9969},
	publisher = {American Physical Society},
	doi = {10.1103/PhysRevB.108.165144}
}

@article{Trama2024TRSWeylSM_SC,
	author = {Trama, Mattia and K{\ifmmode\ddot{o}\else\"{o}\fi}nye, Viktor and Fulga, Ion Cosma and van den Brink, Jeroen},
	title = {{Self-consistent surface superconductivity in time-reversal symmetric Weyl semimetals}},
	journal = {Phys. Rev. B},
	volume = {112},
	number = {6},
	pages = {064514},
	year = {2025},
	month = aug,
	publisher = {American Physical Society},
	doi = {10.1103/bdtb-mb8c}
}

@article{Waje2025PtBi2GL,
	author = {Waje, Harald and Jakubczyk, Fabian and van den Brink, Jeroen and Timm, Carsten},
	title = {{Ginzburg-Landau theory for unconventional surface superconductivity in ${\mathrm{PtBi}}_{2}$}},
	journal = {Phys. Rev. B},
	volume = {112},
	number = {14},
	pages = {144519},
	year = {2025},
	month = oct,
	publisher = {American Physical Society},
	doi = {10.1103/kkqg-ntcz}
}

@article{Bai2025WSMSC,
	author = {Bai, Xuesong and LiMing, W. and Zhou, Tao},
	title = {{Superconductivity in Weyl semimetals with time reversal symmetry}},
	journal = {New J. Phys.},
	volume = {27},
	number = {1},
	pages = {013003},
	year = {2025},
	month = jan,
	issn = {1367-2630},
	publisher = {IOP Publishing},
	doi = {10.1088/1367-2630/ada574}
}

@article{Huang2025WeylIISC,
	author = {Huang, Junkang and Wang, Z. D. and Zhou, Tao},
	title = {{Higher-order topological superconductivity in type-II time-reversal symmetric Weyl semimetals with a hybrid pairing}},
	journal = {Phys. Rev. B},
	volume = {113},
	number = {5},
	pages = {054523},
	year = {2026},
	month = feb,
	publisher = {American Physical Society},
	doi = {10.1103/xy6g-q2f5}
}

@misc{suppl,
    note = "{See Supplemental Material at [URL to be inserted by publisher].}"
}

\end{document}